# High aspect ratio submicron channels using wet etching. Application to the dynamics of red blood cell transiting through biomimetic splenic slits

*Priya Gambhire, Scott Atwell,† Cécile Iss,† Frédéric Bedu, Igor Ozerov, Catherine Badens, Emmanuèle Helfer, Annie Viallat,\* and Anne Charrier\**

*Nanoparticles delivering drugs, disseminating cancer cells, and red blood cells (RBCs) during splenic filtration must deform and pass through the submicron and high aspect ratio interstices between the endothelial cells lining blood vessels. The dynamics of passage of particles/cells through these slit-like interstices remain poorly understood because the in-vitro reproduction of slits with physiological dimensions in devices compatible with optical microscopy observations requires expensive technologies. Here, novel microfluidic PDMS devices containing high aspect ratio slits with submicron width were molded on silicon masters using simple, inexpensive, and highly flexible method combining standard UV lithography and anisotropic wet etching. These devices enabled revealing novel modes of deformations of healthy and diseased RBCs squeezing through splenic-like slits (0.6-2 µm × 5-10 µm × 1.6-11 µm) under physiological interstitial pressures. At the slit exit, the cytoskeleton of spherocytic RBCs seemed to be detached from the lipid membrane whereas RBCs from healthy donors and patients with sickle cell disease exhibited peculiar tips at their front. These tips disappeared much slower in patients' cells, allowing estimating a threefold increase in RBC cytoplasmic viscosity in sickle cell disease. Measurements of time and rate of RBC sequestration in the slits allowed quantifying the massive trapping of spherocytic RBCs.*

Dr. L. Gambhire, Dr. S. Atwell, Dr. C. Iss, MSc. F. Bedu, Dr. I. Ozerov, Dr. E. Helfer, Dr. A. Viallat, Dr. A. Charrier,  Aix Marseille Univ., CNRS, CINAM, Marseille, France
E-mail: charrier@cinam.univ-mrs.fr; viallat@cinam.univ-mrs.fr
Dr. C. Badens, Département de Génétique Médicale, Centre de référence thalassemie, La Timone, Assistance Publique des Hôpitaux de Marseille, Marseille, France. Aix Marseille Univ., INSERM, GMGF, Marseille, France.
† Co-second authors
\*Co-last authors

# 1 Introduction

Vascular endothelial cells line the interior surface of blood vessels. Endothelial cells may have inherent gaps between them, called fenestrations, which present high aspect ratios with submicron widths (from 0.1 to 1 $\mu m$). These fenestrations are crucial in controlling the passage of molecules and cells between the vessels and their surrounding tissues. For example, the large intercellular openings (1.5 $\mu m$) observed in tumors blood vessels may act as a conduit for the passage of tumor cells into the circulation.[1] Conversely, this leaky tumor vasculature causes a differential interstitial pressure,[2] allowing nanoparticles used for drug delivery to migrate from the bloodstream and to accumulate in tumors. The archetypal fenestrated endothelium is found in the venous sinuses of the red pulp of the spleen and plays a crucial role in the mechanical filtering function of the spleen. It is characterized by high aspect ratio interendothelial slits (IESs), the narrowest circulatory bottlenecks in the vascular system, with submicron width (0.2-0.8 $\mu m$), and micron-sized length (2-3 $\mu m$) and height (3-5 $\mu m$) (Figure 1).[3,4] The red blood cells (RBCs) entering the red pulp are forced to squeeze through IESs. As they are disk-shaped capsules of about 8 $\mu m$ in diameter and 1-2 $\mu m$ in thickness (volume about 80-100 $\mu m^3$), they experience extreme deformations within the slits. The less deformable ones, such as aged or long-stored RBCs, get trapped in IESs where they are further eliminated from the blood circulation by white blood cell scavengers. The passage through IESs is also crucial for nanoparticles carrying drugs that must not be cleared out before arriving in tumors. Although in this process, particles/cells size, deformability and shape are believed to be critical parameters playing pivotal roles in the dynamics of the particle/cell interendothelial passage through the epithelium, their effective roles remain poorly understood because in-vivo experiments and observations are often too invasive.[4] This underlines the need for in-vitro observations to improve blood filtration and tumor targeting and to quantitatively understand the dynamics of particles/cells circulating through interendothelial fenestrations. To date such type of device that combines slits with physiologically relevant submicron widths and transparency for the dynamic observation of cells by optical microscopy has not been developed and is lacking.[5]

The first challenge we tackle in this work is to propose an easy, low cost, and highly flexible method to prepare novel microfluidic PDMS (polydimethylsiloxane) devices that integrate high aspect ratio slits with sub-micron width of physiological dimensions and transparency. The difficulty arises from the required combination of high aspect ratio and submicron width of the slits. Typically, their walls must keep a high degree of parallelism to maintain a separation of 0.5 $\mu m$ over a height of 5-10 $\mu m$. PDMS devices are obtained using soft microlithography as replica from a silicon master mold as currently used in microfluidics. Silicon master with submicron dimensions are usually fabricated using electronic lithography followed by deep reactive ion etching (DRIE), a possible method which is costly and slow, and equipment are rarely available in bio-labs. Furthermore, DRIE produces walls surfaces with a relatively high roughness.[6,7] Here we propose an original and ingenious solution which combines standard UV lithography and anisotropic wet etching (AWE).[8] While UV lithography, commonly associated to RIE or DRIE, is limited to the fabrication of features with dimensions larger than 2 $\mu m$, we show that associated to AWE it allows the decrease of structure dimensions to the submicron range with unlimited reduction. Choosing the appropriate silicon wafer orientation and mask designs for the UV lithography we show that a large range of submicron slits can be obtained with width, height and length that can be tuned independently of each other. The slits are perfectly vertical with physiological dimensions (Figure 1C).

The second challenge of this work is to carry out the first in-vitro microscopic observation of the dynamics of RBCs transiting through submicron biomimetic splenic slits in health and disease. In diseases such as hereditary spherocytosis, thalassemia, malaria and sickle cell anemia, RBC deformability is decreased and the splenic circulation, where RBCs have to pass the most stringent 'fitness' test of the microcirculation, is altered. This is a paramount public health issue that causes severe hemolytic anemia and leads to an accelerated splenic destruction of RBCs. Splenomegaly (enlargement of spleen), is observed in hereditary spherocytosis,[9] while splenic infarctions occur in malaria and in sickle cell crisis,[10] which results in a non-functional spleen in the latter case of sickle cell disease. Therefore, recent efforts have been made to address the question of the RBC passage through splenic slits. On the one hand, numerical methods have been developed to describe the passage of RBCs through 1 to 2 $\mu m$ width constrictions.[11,12] On the other hand several experimental studies have proposed in-vitro devices that integrated IES-like slits to observe the dynamics of RBC passage through them, but the width of these artificial slits was never narrower than 2 $\mu m$, i.e. significantly wider than the physiological ones. Two examples are the 'splenon-on-chip' proposed by Rigat-Brugarolas et al.[5] and the series of filtering pillars developed by Picot et al..[13] In these works, the RBCs within the slits did not display the typical dumbbell shape observed in human spleen by electronic microscopy,[14] probably because of the large non-physiological width of the slits. In 2011, Deplaine et al.[4] studied RBC retention in a different device consisting of a layer of microbeads of diameters ranging from 5 to 25 $\mu m$ wherein the interstitial spaces within the beads mimicked IESs, allowing stronger deformation and the observation of dumbbell shapes. However, they did not control interstitial dimensions nor visualize the dynamics of RBC passage. Here, we directly observe and analyze the passage of RBCs through IES-like slits of physiological dimensions and we reveal novel modes of RBC deformation within the slits. These new modes have been made observable due to the strong 3D RBC confinement in our biomimetic slits. Additionally, we show how the alteration of the RBC mechanical properties in two genetic diseases, namely hereditary spherocytosis and sickle cell disease, affects their dynamics of passage through the slits. Our device enables the quantitative characterization of associated shape alterations and of cell retention within the slits of diseased RBCs.

## 2 Results

### 2.1 Device fabrication and mask design

The microfluidic system consists of a main micro-channel (300 $\mu m$ wide and 6 $mm$ long) connected to an inlet, an outlet and two lateral channels (used for rinsing the device) and of internal submicron slits arranged in rows of 5-15 located inside the main channel (Figure 2). Several rows of slits can be placed at successive positions along this channel (Figure S1). The device is obtained using soft lithography by molding PDMS on a silicon master (Figure 2C & D). The silicon master, which is therefore the reverse of the final device (the master must contain micro-bridges in order to get PDMS micro-slits), is fabricated using an audacious combination of UV lithography and AWE which allows the fabrication of structures with submicron dimensions and vertical walls. AWE consists in using the differences in etching rates of the different crystalline planes of silicon when exposed to base solutions (e.g. potassium hydroxide or tetramethylammonium hydroxide) to etch specific crystalline planes[8,15] of the master, obtained after UV lithography, and get structures/bridges with sub-micron dimensions (see details in section Materials & Methods & Figure S3). After AWE, the resulting micro-bridges at the etched silicon surface have therefore shapes and dimensions that differ from the initial mask design (Figure 3i & ii). These changes must be anticipated when designing the mask: for example, micro-bridges with width and length of 11×16 $\mu m^2$ on the mask design (Figure 3i) are reduced to 0.8×2 $\mu m^2$ after AWE (Figure 3.ii).

In the present work we use (110) silicon wafers. These present the advantage of having {111} planes perpendicular to the wafer surface which are used together with the {110} planes to form vertical walls (Figure S2). The initial design leading to a slit is a pair of hexagons carefully oriented with respect to the silicon crystalline planes orientation (Figure 3.i). They are defined by three parameters *a*, *b* and *c*, which correspond to the two lateral dimensions of the mask features (*a* and *b* are respectively parallel and perpendicular to the Si {110} crystalline planes (Figure S2)), and to the distance between the features (c) (Figure 3.i). These parameters are related to the final dimensions of the submicron slits, i.e. the width *w*, the length *l* and the height *d* as shown in Figure 3.ii and must therefore be chosen accordingly. The slit walls (*a*) are parallel to the Si {110} crystalline plane and the four other sides of the hexagonal design have been chosen to have an angle of 35.26° to the wafer flat such that the vertical planes (perpendicular to the silicon surface) along these directions are the {111} planes. These have very slow etch rate and thus play the role of etching stop (Figure S2). The selective etching of {110} lateral planes therefore leads to an extension of the four hexagonal sides and a reduction of the slit walls *a*, yielding a final structure with length $l = a - 2 \times (d/tan(35.26))$ (Figure 3.ii) and width $w = c - 2d$. The height *d* results from the etching of the {110} plane parallel to the silicon surface and is determined by the etching parameters such as time, base concentration and temperature. Cross-sections of the features along the [110] direction and along its perpendicular are represented in Figure 3.iii to illustrate the expected resulting wall verticality and height.

Using the aforementioned relationships and varying the various parameters (*a*, *b*, *c* and *d*), a variety of masters were obtained. Figure 3(A-I) shows some examples of the masters fabricated using our hexagonal design in which the dimensions of the sub-micro-bridges were varied in ranges compatible with the physiological dimensions of splenic slits. The width was varied between 0.6 and 2 $\mu m$, the length between 1.6 and 11 $\mu m$ and the height between 5 and 10 $\mu m$. For each micro-bridge the corresponding mask parameters are reported in Table S1.The dimensions of every slit of all silicon masters were determined by scanning electron microscopy (SEM) (Figure S5). We consider that the PDMS channels obtained after casting on a given master have retained the master micro-bridges dimensions. The small standard deviations of the width and length of the micro-bridges present on the same device (typically ±0.1 $\mu m$), show that the proposed procedure is robust and reproducible. The micro-bridges walls of the silicon masters are vertical and smooth, as checked by scanning electron microscopy (Figure S4(A-C)). Similarly the verticality of the final PDMS slits has been verified by optical microscopy after slicing the device in a direction perpendicular to the length of the main channel across the internal slits (Figure 1C and Figure S4(D)).

## 2.2 RBCs flowing through the submicron biomimetic IESs

Experiments on RBCs passing through the slits have been performed on four different devices, each device containing different slit dimensions. In all devices, both channel and slit heights are the same. D1, D2 and D3 have a height of 10 $\mu m$ and D4 has a height of 5 $\mu m$. The first device, D1, has slits of 2.11±0.10 $\mu m$ in width and 3.45±0.05 $\mu m$ in length, for a volume of 69 $\mu m^3$. Biconcave RBCs (disk of diameter 8 $\mu m$ and thickness 2 $\mu m$) that arrive with an adequate orientation at the slit entrance, represented by both roll and pitch angles equal to zero (Figure 4A), can theoretically flow through the slits without undergoing shape deformation, as indeed observed (Figure 4B). The second and third devices, D2 and D3, have a slit width of ~1 $\mu m$ (Table 1). They mainly differ by their slits length, 2.83 ±0.05 $\mu m$ and 2.26±0.05 $\mu m$, respectively, for a volume of 28.3 $\mu m^3$ and 22.6 $\mu m^3$, respectively. When both roll and pitch angles of the RBC at the slit entrance are equal to zero, the cell is not constrained in height inside the slit but still needs to deform in width because the slit width is narrower than the biconcave disk thickness. The fourth device, D4, has slits dimensions typical of IES ones: 5 $\mu m$ in height, 0.83±0.03 $\mu m$ in width and 1.93±0.03 $\mu m$ in length, for a volume of 8.1 $\mu m3$. Given these dimensions, roll and pitch angles of RBCs within the main channel cannot be both equal to zero and RBCs must therefore strongly deform in the three dimensions to pass through the slits (Figure 4C).

RBCs were obtained from 4 healthy donors. RBCs of one specific donor, obtained from different blood samplings, were studied in three devices, including the one with physiological dimensions. Three pressure drops between the inlet and the outlet of the devices ($\Delta P$) were applied: 5, 10 and 25 *mbar*. The pressure drop per micron ($\Delta P/\mu m$) within the slit (zone 3 in Figure S6) and the average $\Delta P/\mu m$ in the region composed of the slit and of the slanted parts upstream and downstream the slit (zones 2 & 3 in Figure S6) were computed using COMSOL® and are given in Table 1 together with the mean fluid velocity at the slit entrance. In the slit, $\Delta P/\mu m$ is not constant near the entrance and the exit; the in-slit $\Delta P/\mu m$ corresponds to the constant value inside the slit. In the slits of D4 two complementary experiments were performed with RBCs from a patient with hereditary spherocytosis (HS) and from a patient with sickle cell disease (SCD). Complete blood count parameters of these patients are given in Table S2.

### 2.2.1 Healthy RBC deformation: effect of the slit dimensions

In this section, we first describe healthy RBCs shape in upstream height flow, then detail their arrival and passage through the slits, where we distinguish different behaviors according to the slit height.

*Upstream flow.* At pressure drops of 5 and 10 *mbar*, for all devices, and at 25 *mbar* for D4, most RBCs flowing in the main channel upstream from the slit present a non-deformed biconcave discocyte shape and have similar orientations (Figure 4A). Their axis of symmetry is in the direction of observation, with a roll angle of 90°, the disk being parallel to the bottom and top walls of the main channel. In 10 *μm*-high slits (D1-D3), at 25 *mbar* pressure drop, the RBC shape in the main channel is generally not a disk. Most RBCs present a 'banana' body orientated orthogonally to the direction of flow with two thin tails at the rear of the cell due to the friction with the top and bottom walls of the device (Movie S10). Lubrication forces pull the cell membrane at the wall vicinity. This shape is similar to the one observed in wide channels of 5 *μm* in height by Abkarian et al.[16]

*Arrival/passage/exit in 10 μm-high slits (Devices D1-D3).* Most biconcave RBCs approaching a slit entrance change their orientation. Their axis of symmetry tends to rotate so that the cells arrive with both roll and pitch angles equal to zero at the slit entrance (Figure 4A, Movie S8-S10). Within the slit, the shape of a RBC mainly depends on the slit length. Inside longer slits (D1 and D2, $l \geq 2.8$ *μm*), and for $\Delta P$ = 5 or 10 *mbar*, RBCs undergo only a slight deformation (Figure 4B & 5A(i-ii)). Inside shorter and thin slits (D3, $l \leq 2.3$ *μm*), RBCs deform more with large bulges at both entrance and exit of the slit. Typical shapes of RBCs exiting the slits are shown in Figure 5B. In D1 ( *w* ≈ 2 *μm*), many cells are not deformed at the exit and slowly rotate to recover their initial orientation (Figure 4B, Movie S8). The few cells that are larger than 2 *μm* are slightly deformed at the slit exit. In particular, these cells seem to thicken at their front and split in two at their back (Figure 5B(i)). Coherently, in D2 and D3 devices which are narrower than the RBCs (*w* ≈ 1 *μm*), the deformation of the cells at the slit exit is larger with a continuous range of deformation, as illustrated in Figure 5B(ii-vi) and Movie S9-S10, which follows that observed in D1 (Figure 5B(i)). Thus the most striking point here is that exit shapes follow a similar pattern whatever the slits dimensions and pressure drops, with a rear part split in two and a general shape that can be comprised in a triangle with more or less concave sides. Both slit length and pressure drop do not appear as determining parameters for the exit cell shape.

*Arrival/passage/exit in 5 μ m high slits (Device D4).* These slits have dimensions typical of IES ones. To reproduce physiological dynamic conditions, inlet and outlet pressures of 5, 10 and 25 *mbar* have been selected and resulted in RBC transit times through the slits in the range of 60 to 700 *ms*, similar to those reported in the literature in in-vivo conditions on rats (median time of 100 *ms*).[17] When approaching the slits, the RBC velocities are of the order of 100, 200 and 550 $\mu m.s^{-1}$ for the respective pressures. The geometrical confinement in height prevents RBCs from changing their orientation at the approach of the slits as illustrated in Figure 4C and Movie S11. They squeeze within the slits and sometimes fold while entering them. When a cell passes through a slit, most of it is out of the slit, as the cell's volume is much larger than that of the submicron slit; The front and rear of RBC are bloated on both side of the slit (Figure 5A(iv)). Under the incoming flow the volume and the membrane of the cell progress through the slit and the rear bulge at the slit entrance gradually reduces. When it disappears, the part of the cell remaining in the slit quickly exits, and the cell forms an asymmetrical concave region at its rear. The cell finally recovers its initial shape in a slower process. Most of the time a dumbbell shape is observed during the cell passage (Figure 5A(iv) & 6A & Movie S1), which is very similar to that reported in the literature for RBCs in IESs, obtained from electronic microscopy. [4,14] Surprisingly, in some cases, a tip reminiscent of the slit shape is formed at the front of the cell (Figure 5A(v) & 6B &

Movie S2). In this case, the shape of exiting RBCs presents both a concavity at its rear and a tip at its front that disappears in less than 100 $ms$ (Figure 5B(iv)). The occurrence of this specific shape is discussed in the next section. In long slits (0.62 × 5.93 × 5 $\mu m^3$), we observed that RBCs remain blocked inside until the pressure drop is increased to 150 $mbar$ (Movie S12). At the exit, the RBCs were clearly damaged and were unable to recover their normal shape, even within 0.5 $sec$ (RBCs transiting though D4 slits relax within a few tens of milliseconds).

These results clearly show that the height of the channel/slit controls the ability of the RBC to reorient just before entering the slit; this reorientation allowing the cell to limit its deformation within the slit. A robust pattern of exiting-cell shape is observed when the height of the slit does not confine the cell in one dimension. However, when the slit dimension induces strong cell deformation in the three dimensions, a novel shape is observed with a tip forming at the cell front.

### 2.2.2 Healthy and sickle RBC deformation: tip formation

RBCs from a donor with sickle cell disease (SCD) were observed during their passage through the slits in D4. The first result is that almost all cells successfully pass through the slits, even the irreversible sickled-shape cells that arrive with a non discocyte shape (Figure 6(D-G), Movie S4-S7). When focusing on SCD RBCs arriving with a discocyte shape, three exiting cell shapes have been observed and classified into shape 'without tip', shape 'with tip' and 'elongated' shape (Figure 7A). The elongated shape resembles irreversible sickle RBC shape. It represents less than 10% of the exiting RBCs arriving with a discocyte shape.

SCD RBCs display a strong increase in the occurrence of the tip compared to healthy ones (Figure 7B-C). Whereas a maximum of 40% of healthy RBCs present a tip whatever the pressure, this ratio increases up to 70% for SCD RBCs at 5 $mbar$. The difference between the two populations is alleviated with increasing pressure such that almost an equal number of cells exit with and without tips at 25 $mbar$.

The lifespan of the tip at the cell exit was measured on healthy and SCD RBCs. This relaxation time is defined as the difference between the initial time when the tip appears while the cell is within the slit and the final time when the tip has vanished. It is displayed at 10 $mbar$ in Figure 7D. It is three fold higher for SCD RBCs than for healthy RBCs (Mann-Whitney test, $p < 10^{-3}$) and much more dispersed. This relaxation time is directly related to the rheological properties of RBCs and is discussed later.

### 2.2.3 Hereditary spherocytosis RBC deformation: A possible membrane/cytoskeleton dissociation

RBCs from a donor with hereditary spherocytosis (HS) were observed during their passage through the slits of D4 at 25 $mbar$. About 50% of HS RBCs get stuck in the slits while the other 50% pass through. In the slit, HS RBCs form a dumbbell shape similarly to healthy cells (Figure 8B). However as they exit the slit, most of them present shapes akin to a 'jellyfish' with a rounded front part with strong contrast and a rear part with a concave portion and a low contrast (Figure 6C, Movie S3). The high contrast which is specific of concentrated hemoglobin (Hb) indicates that Hb is located at the front part whereas the low contrasted rear part is merely only membrane. It strongly suggests that the two envelopes of the RBC membrane, the lipid bilayer and the cytoskeleton, are dissociated in the rear half of the RBC. The spherical cytoskeletal envelope seems to constrain Hb while the free membrane is stretched to the rear during exit and relaxes only later. To our knowledge, it is the first time that such shapes are observed on flowing RBCs.

### 2.2.4 RBC entrapment

As stressed above many HS RBCs get stuck in the slits. The entrapment of healthy, SCD and HS RBCs were studied at the three pressure drops. An example of entrapment is shown on HS RBCs in Movie S3: while the two first cells pass through, the third one remains blocked in the channel for at least 10 sec. In the present study, a RBC was considered as entrapped when it remained trapped in the slit for longer than 1 second; the others were classified as transiting. This sequestration time was chosen for convenience, as it corresponded to at least ten times the mean transit time we observed for a RBC passing through a slit; it is not related in any manner to the time of survivability of RBCs in the spleen.

The percentage of entrapped and transiting cells for healthy, SCD and HS RBCs is reported in Figure 8. Whereas more than 80% of healthy and SCD RBCs transit through the slits at all three pressures, the entrapment ratio drastically increases in the case of HS RBCs (Figure 8C). At 5 *mbar*, only 13% of HS RBCs transit through the slits; the ratio increases to 50% at 10 and 25 *mbar*. HS RBCs thus have substantially higher entrapment percentage when compared to healthy and SCD RBCs.

## 3 Discussion

We have developed a robust method, which combines standard UV lithography and anisotropic wet etching with soft lithography to obtain series of micron and sub-micron slits in a microfluidic channel. The slits dimensions within the different series were homogeneous, with vertical parallel walls and low roughness and were varied independently in physiological ranges.

We showed that the dynamics of passage of RBCs through the slits could be successfully observed and analyzed. In particular, we revealed that a discocyte RBC is able to reorient its position when approaching a slit of 10 *μm* high and we discovered a generic pattern of cell shape deformations at the slit exit. Interestingly, this pattern is significantly different from the ones predicted by two numerical studies performed with similar slit dimensions and transit times.[12,18] The origin of this discrepancy has to be clarified. It might come from the stress-free shape of RBCs, whose role has not been investigated in these computational studies. The RBC stress-free shape, which is usually considered to vary from a spherical shape to a biconcave disc, is still a subject of debate. It has been shown to play an important role in the motion of RBCs flowing in shear flow.[19] At the approach of the smallest slits of submicron widths and 5 $\mu m$ height, RBCs cannot reorient themselves and must squeeze through the slits.

In our experiments with IES-like slits a novel shape is observed at the slit exit that has not been previously reported in experimental and computational literatures, where a tip is present at the cell front. For instance, the formation of a dumbbell shape within the slit and the existence of a concave rear portion at the slit exit were predicted in a simulation combining a fluid-cell interaction model with a multiscale structural model of the cell membrane.[12] But in this study as well as in the work of Freund et al.[18] the front of the cell remains round, even when the cell folds inside the slit, independently of its initial orientation. The persistence of this tip over tens of milliseconds while the rest of the exiting cell does not show such a deformation raises a question: why does the tip deformation of the cell front relax slower than the other parts of the cell, although every cell part adopts the slit shape when it passes through? A detailed analysis by numerical tools of the forces that apply to the exiting cell, and in particular, to the cell front would clarify this point. Furthermore, tip relaxation and shape recovery are dependent on the viscous dissipation within the membrane and in the hemoglobin (Hb) solution. Although the quantitative determination of the mechanical parameters of RBCs from their tip lifetimes is beyond the scope of this paper, the differences in tip lifetimes between healthy and SCD cells can be used to estimate differences incell viscosity. Both membrane and cytoplasmic viscosity influence the recovery time of RBCs at the slit exit.[20] The membrane viscosity of healthy and HbSS RBCs were studied by G. B. Nash et al..[21] The authors performed a comprehensive set of micropipette experiments and showed that the membrane viscosity of HbSS RBCs with a discoid shape did not significantly differ from that of healthy RBCs. The membrane viscosity was calculated from the product of the shear modulus and the time constant for shape recovery after release of extended RBCs. These results suggest that the increase of relaxation time is mainly due to the increase of cytoplasmic viscosity. Indeed, in SCD RBCs, part of the Hb is mutated in so-called HbS, which form stable polymers in absence of oxygen. It has been suggested that HbS polymers/oligomers could be present in SCD RBCs even in oxygenated conditions and could be responsible for the observed larger dispersion and, on average, a threefold increase of their cytoplasmic viscosity compared to that of healthy RBCs.[22] Here, we also observe a large dispersion and threefold increase in the average tip lifetimes between healthy and SCD cells. Our results thus strongly support those of Byun et al.[22] and show that specific mechanical parameters of the RBCs can be derived from their dynamics of passage through the slits. In the case of healthy RBCs, surprisingly, 30% of the cells form a tip though they do not contain mutated Hb. However, blood samples contain RBCs of varying age and it is well known that old RBCs have an increased cytoplasmic viscosity.[23] We therefore believe that the quantification of tip formation could therefore discriminate the population of the oldest cells in the blood of a donor.

The shape deformations observed on HS RBCs, with a disruption between the lipid bilayer and the cytoskeleton, are very specific of this disease. In hereditary spherocytosis, the anchoring between the lipid bilayer and the underlying 2D cytoskeleton, which together constitute the double envelope of the RBC membrane, is weaker than in healthy cells. Furthermore, this altered anchoring is responsible for a progressive loss of lipids of the membrane during cell aging,[24] resulting in RBCs with a more spherical shape and a decreased deformability. In this context, our results can be interestingly discussed in the light of the numerical simulation by Salehyar and Zhu evoked previously.[12] The simulation predicts the formation of two membrane tongues at the rear of the exiting RBCs, a shape which was never observed in our experiments. However, an interesting point is the computing of the 'contact pressure' between the cytoskeleton and the lipid bilayer. This contact pressure is the contact force per unit area that applies between the two envelopes of the RBC membrane (due to acting of transmembrane proteins that link the two layers). A negative contact pressure corresponds to a dissociation trend between the two envelopes. Salehyar and Zhu showed that the contact pressure in the membrane displays strong negative values at the rear of RBCs while exiting the slit.[12] In HS RBCs, the weaker coupling between the two layers, resulting from defects in transmembrane proteins, combined to the high dissociation force exerted during the passage through the slit may dissociate the two envelopes, thus explaining the jellyfish shapes observed on HS RBCs.

Concerning the higher entrapment of HS RBCs, it is known that these RBCs are characterized by a reduced deformability and a more spherical shape, and display a reduced life span and a strong splenic entrapment. However, the direct link between their mechanical and morphological properties and their splenic sequestration is little documented. Recently, a direct link between the RBC surface area-volume ratio (a marker of both morphology and deformability) and the extent of splenic entrapment was evidenced by perfusing ex vivo normal human spleen with RBCs displaying various degrees of surface area loss.[25] The experimental data, however, obtained from ektacytometric and flow cytometric analyses, and independently performed on large heterogeneous populations of RBCs, did not allow the direct individual measurements of both the morphology and the passage through IESs of a given cell. Therefore, some of the results, such as the fact that a subset of RBCs with a high sphericity index managed to transit through the spleen, were difficult to interpret. In particular, the scenario suggested in this study, i.e. that a RBC stuck in an IES could lose some of its volume in order to pass through the slit would have required individual cell observation and was therefore not tested. Our devices enable the visualization of the dynamics and retention of individual RBCs through IES-like slits and make therefore possible the direct correlation between cell morphology and entrapment. Indeed, it is the first time that the entrapment ratio during a given time is quantitatively measured in-vitro by visualizing trapped cells. In addition, it allows quantitative comparison of the retention of healthy and sick RBCs in the same IES-like slits. Finally, the possible membrane loss by vesiculation, volume leak, due to increased permeability or ion exchange by mechanosensitive ionic channels could be directly detected on single entrapped cells, thus enabling the study of the active mechanisms of cell deformation under a strong confinement.

## 4 Conclusions

In summary, the present study reports the first fabrication of micro-channels containing high aspect ratio submicron slits of physiological relevance for endothelial fenestrations. These channels enabled the full observation, under different pressure drops, of the dynamics of individual RBCs undergoing strong deformations when transiting through the slits. Novel shape deformations have been discovered that are not predicted by numerical simulations, thus posing a numerical and conceptual challenge. Moreover, specific behaviors of pathological RBCs were highlighted, such as the cell entrapment and the dissociation of the cell membrane envelopes for RBCs from patients with hereditary spherocytosis (HS) and the prolonged presence of a tip at the front of RBCs from patients with sickle cell disease (SCD). Both phenomena, reported for the first time in in-vitro submicron fenestrations, definitely underline the importance of strong cell confinement and deformation. These specific behaviors shed light on the alteration of the mechanical properties of RBCs in disease (weak coupling between the lipid bilayer and the cytoskeleton for HS and increased cytoplasmic viscosity for SCD). A forthcoming challenge will be to couple experiments with computations to determine the mechanical properties of flowing RBCs from their dynamics through the slits. These devices can also be used to study the dynamic behavior of RBCs trapped into the slits, such as a volume change induced by the activation of a mechanosensitive channel, for instance PIEZO1. We definitely believe that the approach presented in this work will enable to address numerous crucial problems in relation to the RBC splenic passage and to key physiological processes involving endothelial fenestrations.

## 5 Materials and methods

### 5.1 Detailed silicon master fabrication protocol

A detailed description of the device fabrication steps is schemed in Figure S3. All lithography procedures have been carried out in the clean room facility PLANETE (CT-PACA Micro- and Nano-fabrication Platform, Marseille, France). The device was fabricated from scratch using a (110) ± 0.5° 4-inch silicon wafer (Siltronix, France) covered with a ∼ 100 $nm$ thick thermal oxide layer (Renatech, France) (Figure S3(1)). Prior to any process the wafer is cleaned using sonication in solvent baths of acetone and isopropanol (IPA) (Sigma Aldrich), consecutively, and exposed to a 300 $W$ oxygen plasma treatment for 10 $min$ in a plasma cleaner (NanoPlas, France). Device channels are obtained by photolithography as depicted in Figure S3(2-4). The wafer is spin-coated at 4000 $rpm$ for 60 $sec$ with positive resist Microposit S1813 (Dow Chemical Company), then pre-baked 60 $sec$ at 115°$C$. After UV exposure through the mask, the resist is developed using developer MF 319. The pattern is then transferred to the silicon oxide layer using reactive ion etching (RIE) (MG200 Plassys, France) with CHF3 gas at 10 $cm^3\ min^{-1}$ with a 70 $W$ RF power. The complete removal of the oxide layer is monitored by a LASER interferometer. After resist removal the wafer is cleaved into approximately 2 × 2.5 $cm^2$ pieces using a diamond wire saw. The final stage of the master mould fabrication consists in transferring the pattern on the silicon oxide layer to the silicon (Figure S3(5)). This is obtained by anisotropic wet etching using 20% KOH solution (Sigma Aldrich) with 10% IPA at 70°C. The etching is stopped by dipping the wafer in a water bath. The remaining oxide layer, is then removed by dipping the wafer in a 10% HF bath for 10 *min* (Figure S3(6)). The resulting mould presents a large structure and several sub-micro-bridges corresponding to the main channel and the sub-micrometric channels respectively.

## 5.2 Microfluidic device preparation

The final PDMS device is obtained by molding on the silicon master as described in Figure S3(7). To prevent the PDMS from adhering to the silicon surface, the latter is first made hydrophobic by functionalizing its surface with tricholroperfluorooctyl silane by chemical vapour deposition. PDMS (10:1, PDMS:curing agent, Sylgard 184, Dow Corning) is poured over the silicon master mould and thawed at 65°C for at least 2 *hr*. After polymerization, the PDMS is peeled off carefully from the master. Access to the inlets and outlets of the microfluidic device in PDMS is created by punching holes using a 0.75 *mm* diameter punch (Harris Unicore). The PDMS device and a glass coverslip (60 × 24 $mm^2$) are finally plasma treated for 1 and 5 *min*, respectively, and bonded together at 65∘C for at least 2 *hr* to obtain the final device (Figure S3(8)).

## 5.3 Red blood cell sample preparation

Blood samples are obtained from a healthy donor, a patient with sickle cell disease (SCD), and a patient with hereditary spherocytosis (HS). The patient specimens are a subset of residual samples referred to the Department of genetics (La Timone Hospital, Marseille, France) for routine tests. The blood is mixed with EDTA (anticoagulant) as soon as harvested, then processed as follows to extract the patient RBCs. Samples are centrifuged at 500 *g* at 4°C for 10 *min* to pellet the RBCs. The pellets are resuspended in Sagmanitol (SAGM) (EFS, France) and washed three times at the same centrifuge settings. After washing, the hematocrit (Hct) is maintained at 50% in SAGM. RBCs samples thus prepared are stored at 4∘C and used within 7 days, i.e. during a period of time short enough to prevent alteration of RBC deformability[26,27]. Blood sample from the healthy subject is obtained by a pinprick (approximately 20 $\mu l$) and immediately added to 500 $\mu l$ of SAGM. It is washed three times at the same centrifuge settings as those used for the patient samples. After washing, Hct is maintained at 50% in SAGM and used within the day. Experiments are performed at 1% Hct, after diluting 3 $\mu l$ of 50%-Hct RBC suspensions in 150 $\mu l$ of phosphaste buffered saline (PBS) with 0.1% bovine serum albumin (BSA) (Sigma Aldrich).

## 5.4 Passage of RBCs through the microfluidic system

The flow of RBCs through the microfluidic device is controlled using a microfluidic flow controller system (MFCS-8C, Fluigent). The connections to the MFCS are made through teflon tubing (0.015 and 0.027 *inch* inner and outer diameters, respectively) inserted into the inlet and outlet pores of the device. The setup is mounted on an inverted microscope (Olympus IX71) equipped with a 100× objective. The images of the flow are obtained in bright-field using a high-speed camera (Fastcam Mini, Photron, Japan) and stored as TIFF stacks. Image processing is carried out using ImageJ.

## Acknowledgement

This work has been carried out thanks to the support of the A*MIDEX project (n◦ ANR-11-IDEX-0001-02) funded by the Investissements d'Avenir French Government program, managed by the French National Research Agency (ANR). The authors would like to acknowledge CT-PACA PLANETE facility for access to the microfabrication tools and the french RENATECH network for the thermal oxide layer deposition on the silicon wafers. We also would like to thanks Nahid Hassanpour for Comsol® simulations.

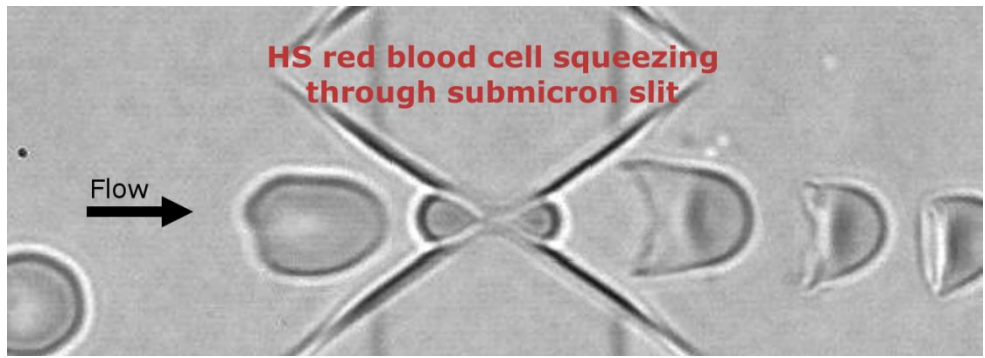

TOC: Microfluidic PDMS devices containing high aspect ratio slits with submicron width are used as biomimetic inter-endothelial spleen slits to study the mechanical properties of red blood cells (RBCs). New deformation shapes are observed with RBCs from patients with Sickle Cell Disease and Hereditary Spherocytosis (HS). For example, HS cells display possible membrane/cytoskeleton dissociation at the slit exit.

| Slit | | Flow velocity (*µm/s*) | | | Average \| In-slit pressure drop (*Pa/µm*) | | |
|---|---|---|---|---|---|---|---|
| | | I/O ΔP (*mbar*) | | | I/O ΔP (*mbar*) | | |
| Type | Dimensions ($µm^3$) | 5 | 10 | 25 | 5 | 10 | 25 |
| D1 | 2.11±0.10 × 3.45±0.05 × 10 | 184 | 368 | 921 | 1.02 \| 12.0 | 2.04 \| 23.9 | 5.10 \| 59.8 |
| D2 | 1.10±0.10 × 2.83±0.05 × 10 | 119 | 238 | 596 | 6.79 \| 58.0 | 13.58 \| 116.0 | 33.96 \| 274.0 |
| D3 | 0.88±0.05 × 2.26±0.05 × 10 | 147 | 294 | 735 | 6.76 \| 66.4 | 13.52 \| 133.0 | 33.82 \| 333.0 |
| D4 | 0.83±0.03 × 1.93±0.03 × 5 | 95 | 189 | 474 | 5.21 \| 55.6 | 10.40 \| 111.0 | 26.04 \| 278.0 |

**Table 1** Flow velocities in the main channel, average and in-slit pressure drops per micron (*ΔP/µm*) in the slits calculated using COMSOL® for devices D1-D4 at imposed inlet/outlet (I/O) pressure drops (*ΔP*) of 5, 10 and 25 *mbar*. The average *ΔP/µm* has been calculated along the dotted line in the blue/yellow region in Figure S6. The in-slit *ΔP/µm* corresponds to the pressure inside the slit in the constant regime (see Figure S6).

**FIGURE CAPTIONS**

**Figure 1** A simplified schematic representation of the filtration function of the spleen. The red blood cells enter through the arterioles and remain in the red pulp region (A) until they can pass through the interendothelial slits to enter the venous sinus (B). (C) A biomimetic sub-micron interendothelial slit (IES) made of PDMS.

**Figure 2** (A-B) Schematics of the designs to obtain the silicon master of the device. (A) Mask design of the entire device for UV lithography. It is composed of a large main channel (∼ 300 *µm* wide and 6 *mm* long) with four inlets/outlets. The circle highlights the position of the submicron channels mimicking the IES. At this stage their design has micron size dimensions. A zoom of these channels is shown in (B). (C-D) Detailed schematic 3D-views of the silicon sub-micro- bridges obtained after anisotropic wet etching of the silicon (C) and of the resulting PDMS sub-micro-channels (D). The 11 × 16 $µm^2$ junctions in (B) give sub-micro-bridges of 0.8 × 2 $µm^2$ in (C) after chemical etching.

**Figure 3** The hexagonal (i) UV mask design transferred onto SiO2/Si substrate by photolithography, of dimensions (***a,b***) and spacing I, and the corresponding feature obtained upon wet chemical etching of silicon (110) wafer: schematic top views (ii), and cross-section profiles along the planes *QQr* and *PPr* (iii) after etching to a depth ***d*** .(A-I) Series of optical micrographs of resulting etched patterns on silicon with dimensions varying from 0.6 to 2 *µm* in width, 1.6 to 10.8 *µm* in length and 5 to 10 *µm* in height. Scale bar: 10 *µm*.

**Figure 4** (A) Schematics (top view) of roll and pitch angles ($\vartheta$ and $\varphi$ respectively) of a healthy RBC in upstream and downstream flow of a 10 *µm* high channel/slit. (B) Time-lapse of a healthy RBC rotating from $\vartheta$ = 90° and $\varphi$ = 0° to both angles equal to 0° as it enters a 2.11 × 3.45 × 10 $µm^3$ slit. The cell undergoes very little deformation before recovering its initial shape and orientation. (C) Time-lapse of a RBC passing through a 0.83 × 1.93 × 5 $µm^3$ slit. The cell cannot rotate and has to deform strongly to pass through the slit with the formation of bulges at both its front and rear. Scale bar is 10 *µm* in B and C.

**Figure 5** (A) Typical images of RBC passage through slits of various dimensions. In wide and/or long slits (i-ii), the cell does not need to deform, or little, to pass through the slit. In smaller (thin and shorter) slits (iii-v), the formation of large bulges is observed at the front and the rear of the cell (dumbbell shape). In some cases (v), a tip reminiscent of the slit shape appears at the cell front. (B) The shapes of RBCs exiting the slits follow a general pattern which cannot be related neither to the applied pressure nor to the slit dimensions. The selected images, chosen independently of the slit dimensions and for various pressure drops (5,10, 25 *mbar*), illustrate the continuous spectrum of shapes which can all be described as a triangle with more or less concave sides, with a rear split in two parts. Scale bars: 10 *µm*.

**Figure 6** Time-lapses of the three types of RBCs passing through $0.83 \times 1.93 \times 5 \ \mu m^3$ slits showing the various shape changes while the RBCs go through and exit from the slits. All experiments were performed at 5 *mbar*, except for the HS sample (C) at 25 *mbar* to prevent high RBC entrapment and slit clogging. (A-B) Healthy RBCs passing through without tip (A) and with a tip formed downstream (B). (C) An HS RBC which is able to pass through the slit showing a 'jellyfish' exit shape. (D-G) SCD RBCs passing through without tip (D), with a tip upon exit (E) or forming an elongated shape (F). An irreversibly sickle cell (ISC) which is already deformed before passing through the slit is shown in (G). Scale bar: 10 $\mu m$.

**Figure 7** (A) Typical images of RBCs exhibiting the three categories of exit shapes. (B-C) Percentages of cells exiting from the slits without tip (yellow), with tip (green), and elongated (blue) for Healthy (B) and Sickle cell anemia (C) blood at the three pressures. *n*: total number of cells observed at each pressure. (D) Comparison of the tip recovery time for H RBCs (closed triangles, *n* = 30) and SCD RBCs (open circles, *n* = 59) at a pressure drop of 10 mbar. ∗∗ : $p<10^{-3}$.

**Figure 8** Percentages of transiting (red) and entrapped (gray) RBCs for (A) Healthy (B) Sickle cell anemia and (C) Hereditary Spherocytosis blood, while passing through $0.83 \times 1.93 \times 5 \ \mu m^3$ slits at different pressures. *n*: total number of cells observed at each pressure.

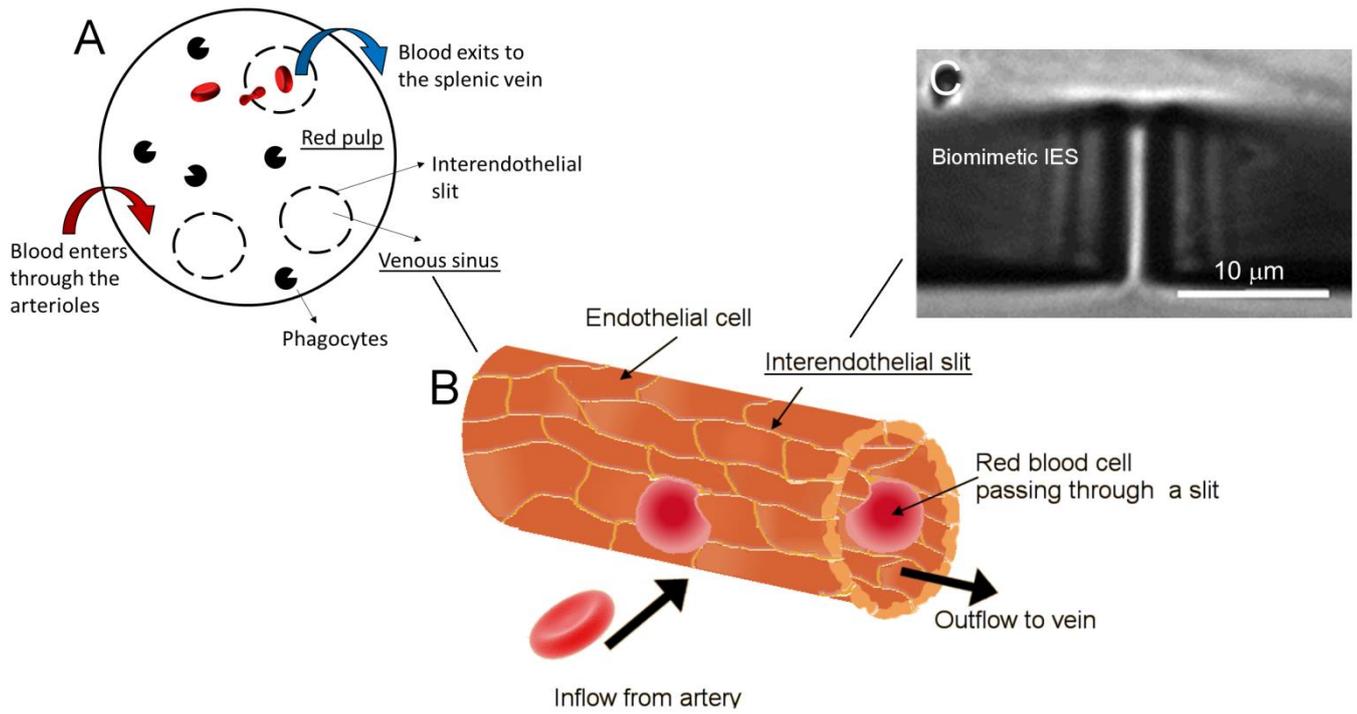

FIGURE 1

(a) 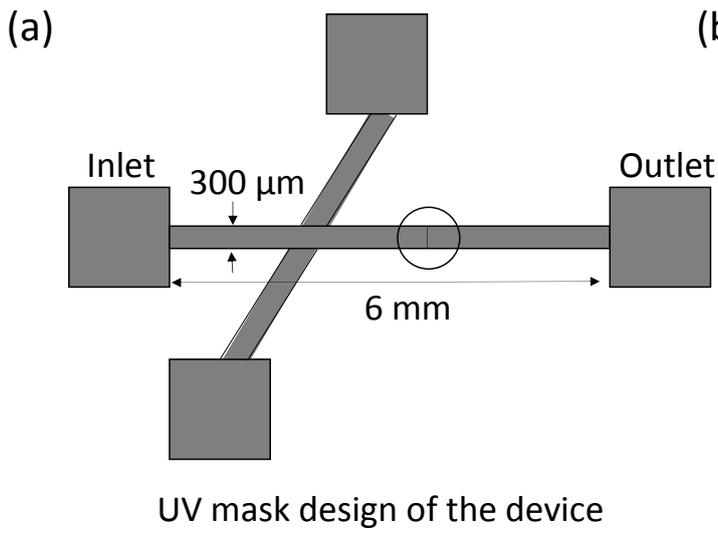

UV mask design of the device

(b) 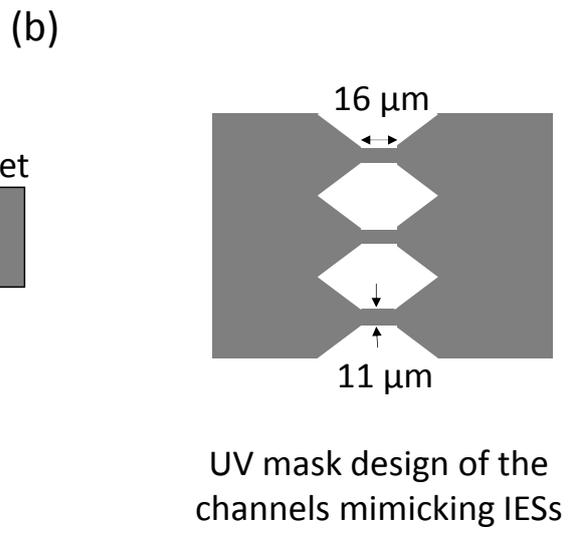

UV mask design of the channels mimicking IESs

(c) 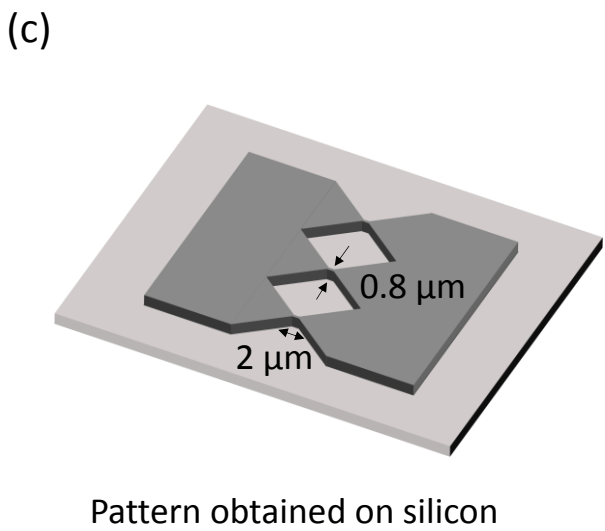

Pattern obtained on silicon

(d) 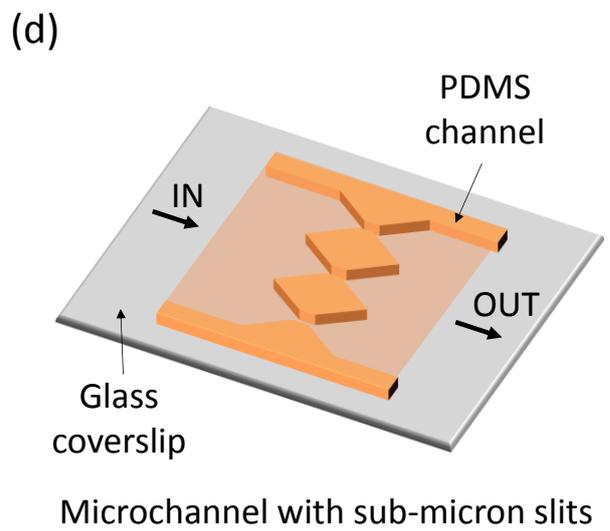

Microchannel with sub-micron slits

FIGURE 2

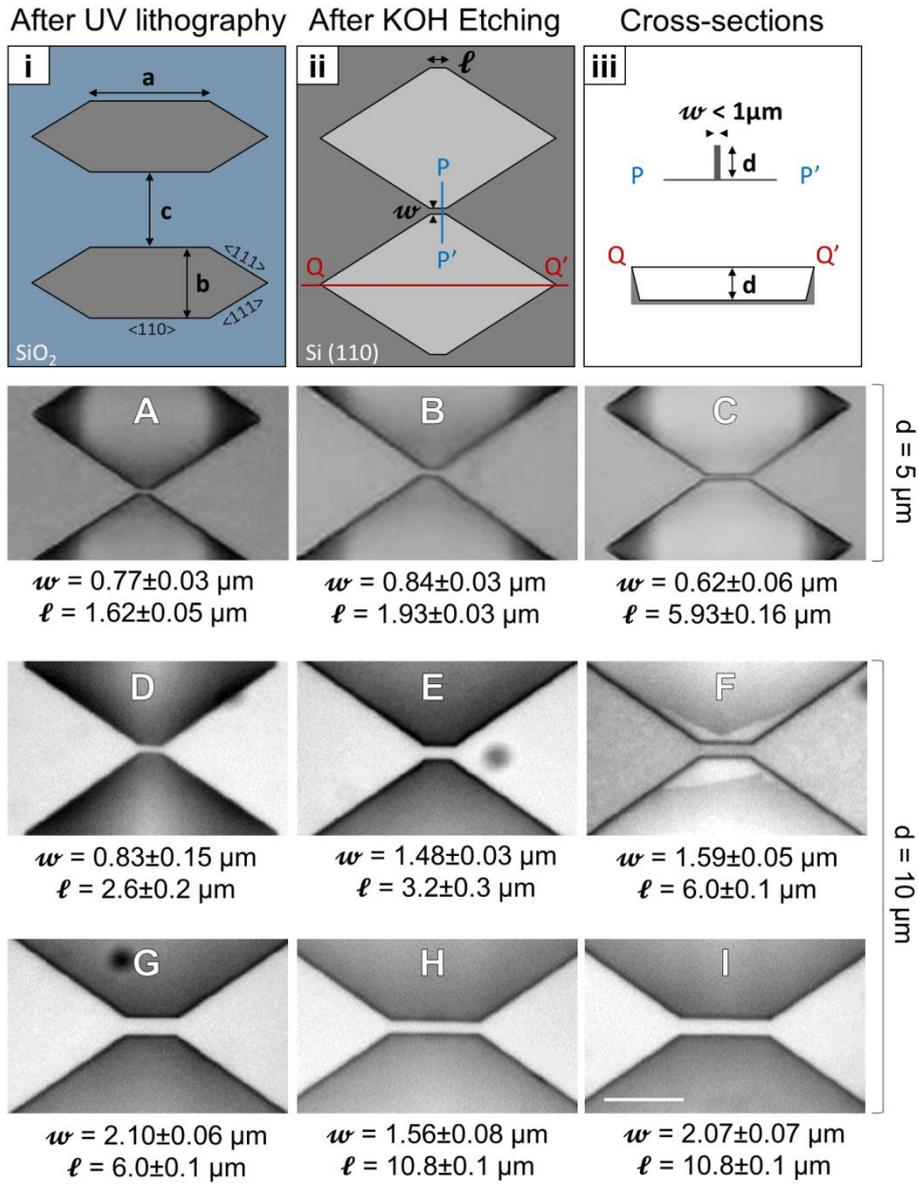

FIGURE 3

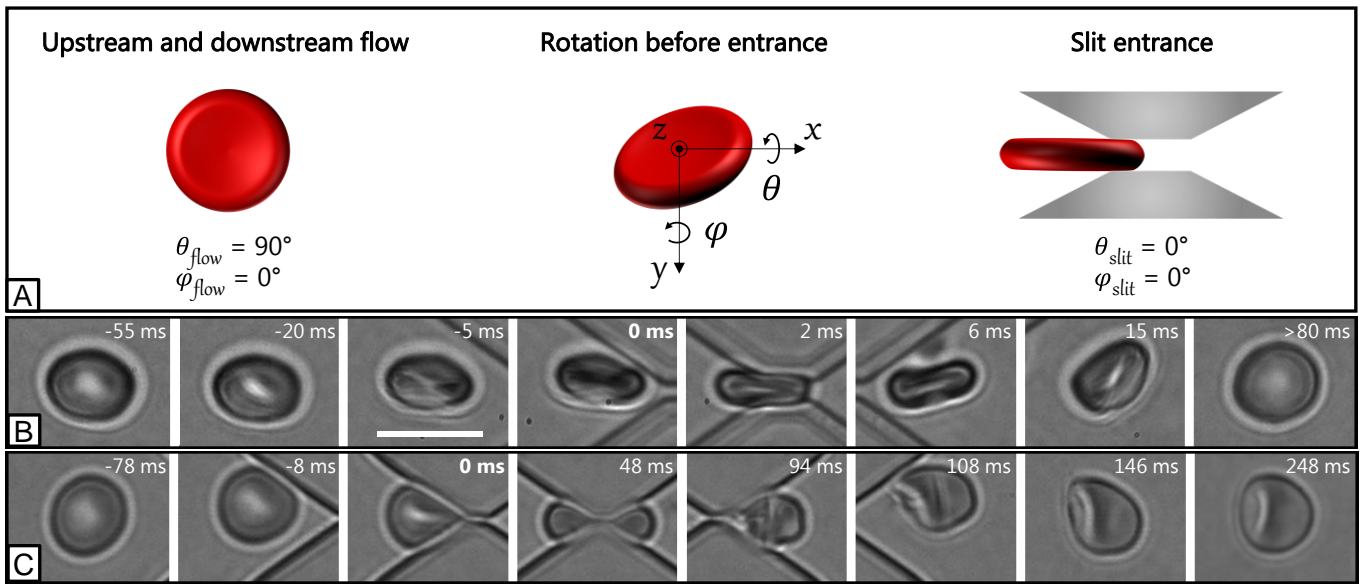

FIGURE 4

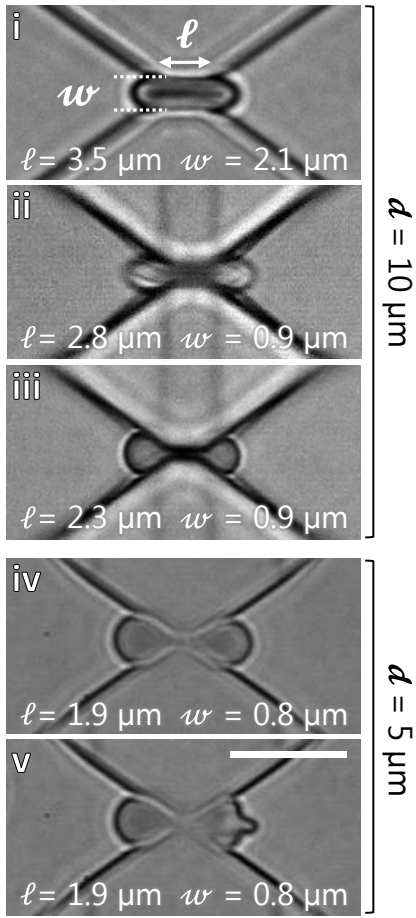
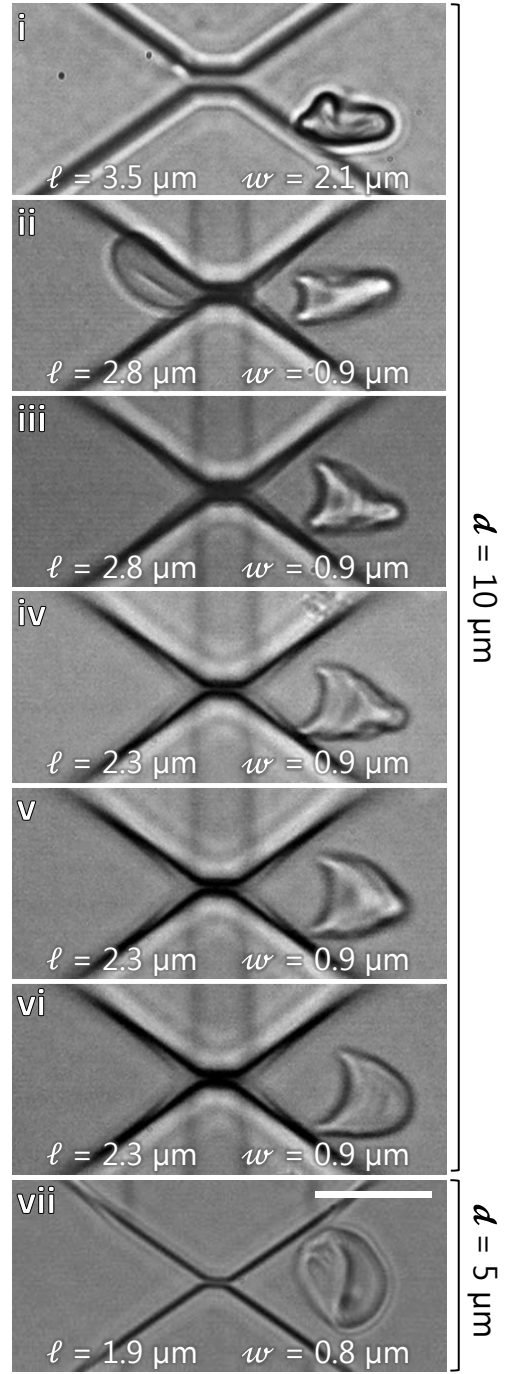

FIGURE 5

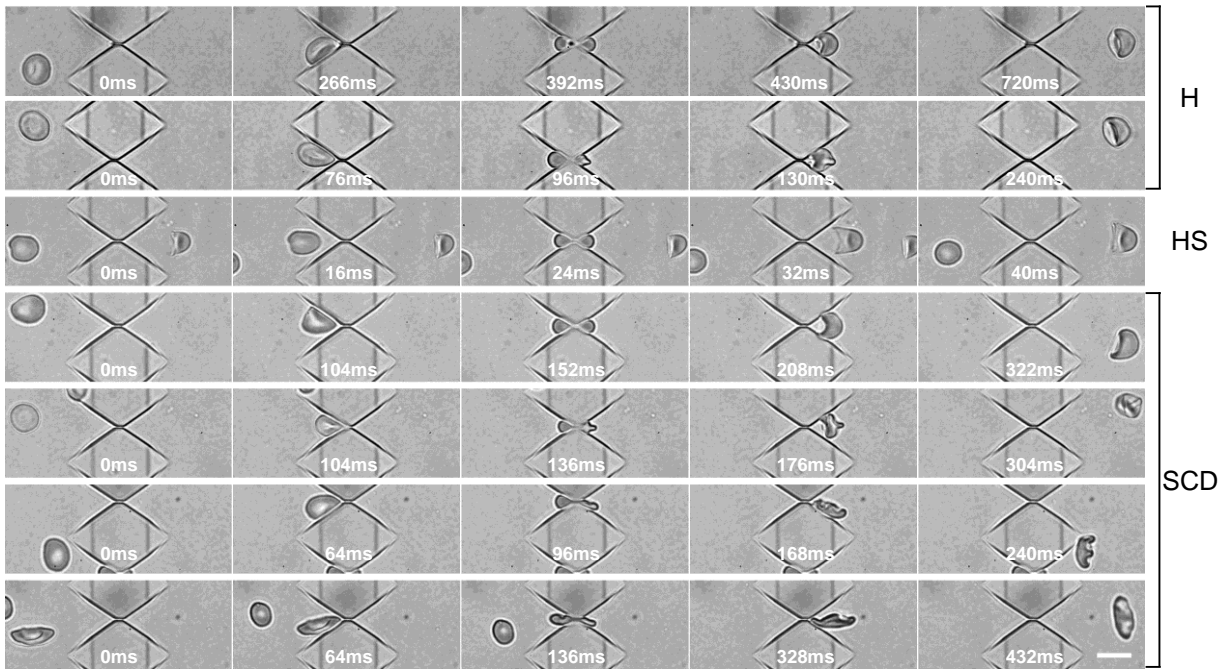

FIGURE 6

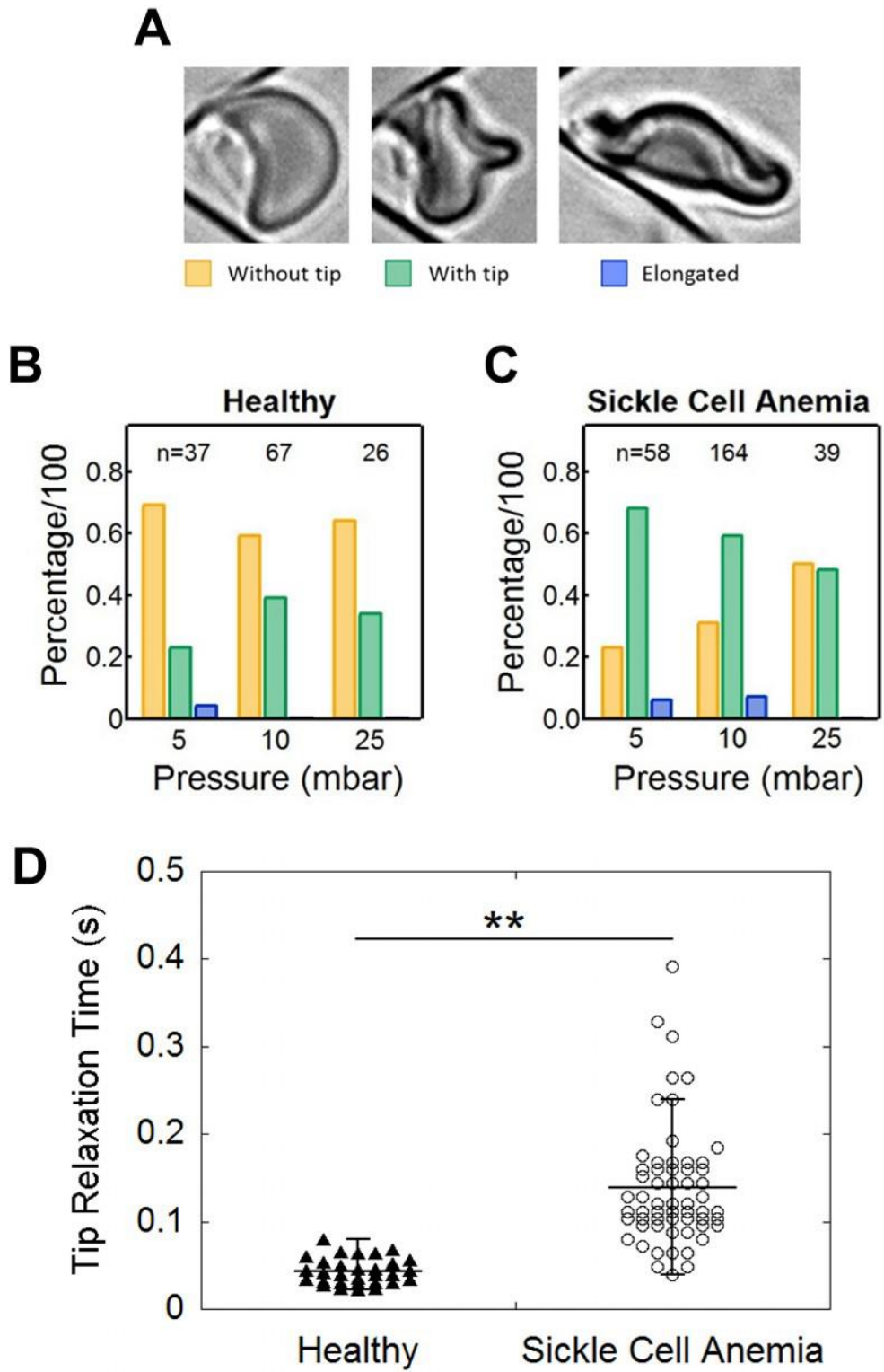

FIGURE 7

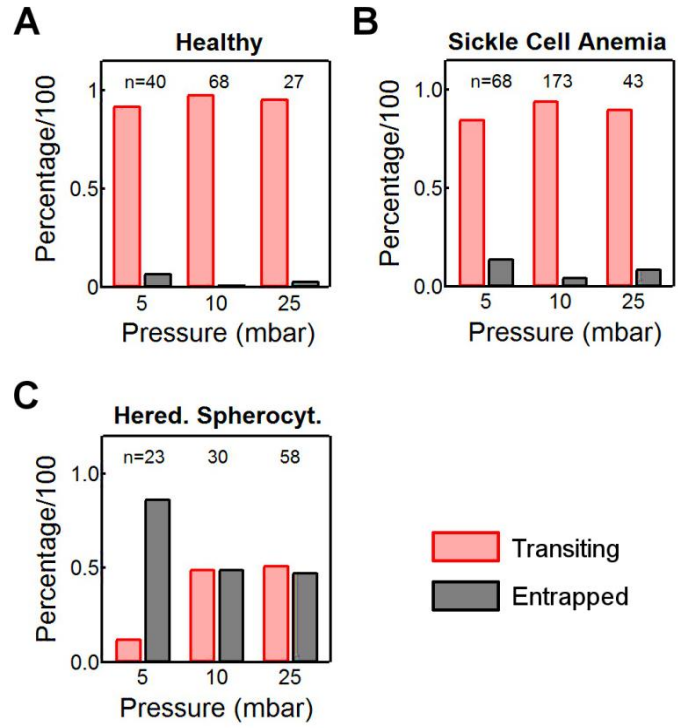

FIGURE 8